\documentclass[sn-mathphys-num]{sn-jnl}% Math and Physical Sciences Numbered Reference Style 

\usepackage{graphicx}%
\usepackage{multirow}%
\usepackage{amsmath,amssymb,amsfonts}%
\usepackage{amsthm}%https://tex.zih.tu-dresden.de/project/65cce192ce7135668164a8ae
\usepackage{mathrsfs}%
\usepackage[title]{appendix}%
\usepackage{xcolor}%
\usepackage{textcomp}%https://tex.zih.tu-dresden.de/project/65cce192ce7135668164a8ae
\usepackage{manyfoot}%
\usepackage{booktabs}%
\usepackage{algorithm}%
\usepackage{algorithmicx}%
\usepackage{algpseudocode}%
\usepackage{listings}%
\usepackage{multicol}
\usepackage{float}
\usepackage[normalem]{ulem}

\theoremstyle{thmstyleone}%
\theoremstyle{thmstyletwo}%

\theoremstyle{thmstylethree}%

\usepackage{upgreek}

\begin{document}

\title[Article Title]{A self-organised liquid reaction container for cellular memory}


\author[1,3]{\fnm{Sukanta} \sur{Mukherjee}}

\author[1,3]{\fnm{Enrico} \sur{Skoruppa}}

\author[2]{\fnm{Holger} \sur{Merlitz}}

\author[1,2,3]{\fnm{Jens-Uwe} \sur{Sommer}}\email{sommer@ipfdd.de}

\author[1,3]{\fnm{Helmut} \sur{Schiessel}}\email{helmut.schiessel@tu-dresden.de}

\affil[1]{\orgdiv{Cluster of Excellence, Physics of Life}, \orgname{TU Dresden}, \orgaddress{\city{01307 Dresden}, \country{Germany}}}

\affil[2]{ \orgname{Leibniz-Institut f\"ur Polymerforschung Dresden}, \orgaddress{\city{01069 Dresden}, \country{Germany}}}

\affil[3]{\orgdiv{Institut f\"ur Theoretische Physik}, \orgname{TU Dresden}, \orgaddress{\city{01062 Dresden}, \country{Germany}}}

\abstract{
Epigenetic inheritance during cell division is essential for preserving cell identity by stabilizing the overall chromatin organisation. Heterochromatin, the condensed and transcriptionally silent fraction of chromatin, is marked by specific epigenetic modifications that are diluted during each cell division. Here we build a physical model, based on the formation of a biomolecular condensate, a liquid ‘droplet’, that promotes the restoration of epigenetic marks. Heterochromatin facilitates the droplet formation via polymer-assisted condensation (PAC). The resulting condensate serves as a reaction chamber to reconstruct the lost epigenetic marks. We incorporate the enzymatic reactions into a particle-based simulation and monitor the progress of the epigenetic markers through an \textit{in silico} analogue of the cell cycle. We demonstrate that the proposed mechanism is robust and stabilizes the heterochromatin domains over many cell generations. This mechanism and variations thereof might be at work for other epigenetic marks as well.
}

%%%%%%%%%%%%%%%%%%%%%%
% FOR SUBMISSION
% Epigenetic inheritance during cell division is essential for preserving cell identity by stabilizing the overall chromatin organisation. Heterochromatin, the condensed and transcriptionally silent fraction of chromatin, is marked by specific epigenetic modifications that are diluted during each cell division. Here we build a physical model, based on the formation of a biomolecular condensate, a liquid ‘droplet’, that promotes the restoration of epigenetic marks. Heterochromatin facilitates the droplet formation via polymer-assisted condensation (PAC). The resulting condensate serves as a reaction chamber to reconstruct the lost epigenetic marks. We incorporate the enzymatic reactions into a particle-based simulation and monitor the progress of the epigenetic markers through an in silico analogue of the cell cycle. We demonstrate that the proposed mechanism is robust and stabilizes the heterochromatin domains over many cell generations. This mechanism and variations thereof might be at work for other epigenetic marks as well.

% \keywords{PAC, Epigenetic inheritance, Heterochromatin, Dynamic, Self-organised}



\maketitle

%%%%%%%%%%%%%%%%%%%%%%%%%%%%%%%%%%%%%%%%%%%%%%%%%%%%%%%%%%%%%%%%%%%%%%%%%%%%%
%%% INTRO
%%%%%%%%%%%%%%%%%%%%%%%%%%%%%%%%%%%%%%%%%%%%%%%%%%%%%%%%%%%%%%%%%%%%%%%%%%%%%
When cells copy their genetic information in preparation for cell division, they face the challenge of also duplicating their epigenetic information. 
Epigenetics refers to stable gene activity regulating modifications beyond the primary
DNA sequence, that are inheritable and persist through cell divisions~\cite{cortini16}.
These modifications add an additional layer of information to the genetic layer and play an important role in defining cell types. At the molecular level, one of the mechanisms by which epigenetic information is stored concerns covalent modifications of histone proteins \cite{cortini16}. 
These
proteins form the core of nucleosomes, the elementary packaging units of chromatin. In a nucleosome, about 150 base pairs (bp) of DNA—roughly its persistence length—are wrapped around an octamer of histone proteins \cite{luger97,schiessel23} and neighboring nucleosomes are connected by short stretches of linker DNA. 
Certain epigenetic marks associated with histone proteins play a vital role in differentiating the morphology of 
% the 
chromatin, i.e., the more open euchromatin or the less accessible heterochromatin. Therefore, cell type-specific marks result in cell type-specific packaging and accessibility of the genetic material.

\begin{figure}
    \centering
    \includegraphics[width=88mm]{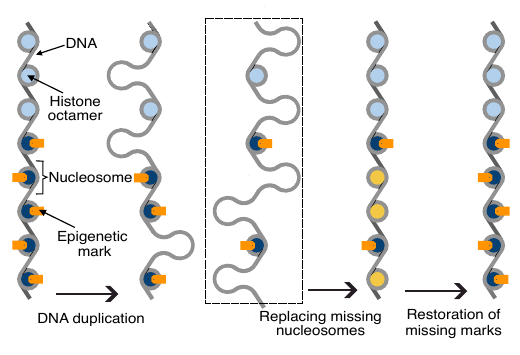}
    \caption{\textbf{Dilution of epigenetic information during cell division:} As a result of DNA duplication the string of nucleosomes along the parent DNA molecule gets distributed between the two daughter DNA molecules. Missing nucleosomes are then replaced by the addition of new histone proteins. The challenge is to re-establish the missing epigenetic marks. The original sequence of nucleosomes contains light blue and dark blue octamers, the latter carrying epigenetic marks (orange). New nucleosomes with missing marks are highlighted in yellow.}
    \label{fig:general_scheme}
\end{figure}

During DNA duplication, the nucleosomes are randomly distributed between the two daughter chromosomes, and vacant nucleosome positions are filled by new and unmodified nucleosomes~\cite{yu18}, see Fig.~\ref{fig:general_scheme}.
Since only the old nucleosomes carry the original epigenetic marks, this process results in a ``dilution'' of marks by a factor of two. 
To prevent the loss epigenetic information, a mechanism to reconstruct the missing marks is required.

Earlier attempts to understand how epigenetic information is established and propagated through 
% the 
cell divisions typically invoke scenarios where epigenetic marks spread along a one-dimensional array of nucleosomes \cite{bannister01,nakayama01,gaszner06}. For example, an important epigenetic mark is H3K9me3, a trimethylation modification on histone H3—one of the four histones that make up the histone octamer. 
Nucleosomes with this mark typically occur in longer blocks of many such nucleosomes (in humans, blocks have a median length of 50 nucleosomes \cite{barkess12}). There is a protein called heterochromatin protein 1 (HP1) that has a specific binding site for this mark \cite{bannister01,nakayama01,lachner01}. This led the authors of Ref.~\cite{bannister01} to propose the following scenario: HP1 binds to nucleosomes with the H3K9-mark which in turn recruits an enzyme called SUV39H1 methylase; when bound to HP1, this enzyme methylates neighboring non-methylated nucleosomes.
As a result, a methylation wave propagates through the one-dimensional chain of nucleosomes up to the point where it encounters a ``boundary element'' \cite{bannister01,gaszner06,barkess12}. However, the physical nature of such boundary elements remains unclear.

The missing element in this one-dimensional picture is the three-dimensional folding of chromosomes.
Different parts of chromosomes might not be as effectively isolated from each other and the methylation wave might jump over to a nucleosome nearby in 3D space but far away along the chain of nucleosomes. Moreover, even if a perfectly one-dimensional mechanism existed, an inadvertently missing nucleosome in a chain of nucleosomes could completely stop the spreading of the methylation marks along the array.

This suggests that it is necessary to consider an alternative mechanism in which the restoration of the methylation marks after DNA duplication relies on the three-dimensional structure of the chromosome. Contact maps from chromosome capture experiments provide the relevant information to this problem. Such maps highlight nucleosomes with an H3K9me3 mark to have a much higher probability of being close to other nucleosomes with this mark than to nucleosomes without the mark \cite{rao14}. 
Contact maps reveal the presence of two primary chromatin compartments: euchromatin, characterized by a relatively open structure with nucleosomes largely lacking the H3K9me3 mark, and heterochromatin, which is more densely packed and primarily composed of H3K9-methylated nucleosomes.

In addition to accounting for the 3D chromatin structure, a physical scenario should satisfy at least the following three requirements: 
(A) The epigenetic state must be preserved across not only a single- but up to about 50 cell divisions, the Hayflick limit \cite{hayflick65}.
(B) Experiments based on quantitative mass spectroscopy indicate that the reconstruction of the marks is
extremely slow, of the order of 20 hours \cite{xu12}. Any mechanism that requires processes to occur on a much shorter time
scale is therefore unlikely to reflect the real system. (C) Candidate mechanisms should not critically depend on particular parameters—i.e., they should not require fine-tuning—as many system characteristics vary with time (see below) and are vastly different across different cell types.

A substantial body of literature already exists exploring various aspects of epigenetics from a physics perspective
\cite{dodd07,micheelsen10,dodd11,zhang14,tian16,erdel16,michieletto16,jost18,michieletto18,sneppen19,sandholtz20,nickels21,katava22,abdulla22,owen23,skjegstad23}. Of those, two simulation studies \cite{sandholtz20,owen23} make explicit use of the three-dimensional structure of chromosomes to reconstruct the marks lost during duplication. However, the first study \cite{sandholtz20} does not meet the requirements listed above. 
In this study, chromatin is modelled as a highly coarse-grained copolymer where heterochromatin regions compact in the presence of HP1. Subsequently, the chromosome conformation is frozen and methylation marks are removed. Still frozen, epigenetic marks are reassigned to individual monomers with probabilities based on the density of HP1 proteins in the vicinity of a given monomer. 
Only after the catalytic process is complete, the polymer is free to move again, and the entire process is repeated.
The system is found to be highly sensitive to parameters, with small parameter changes leading to large changes in overall methylation levels, and thus does not satisfy requirement C. Methylation patterns are maintained for only a few cell generations under optimal parameters and thus do not meet requirement A. Remarkably, this occurs despite freezing of the chromosome during methylation reactions. This assumption of instantaneous reactions is not consistent with requirement B.

The more recent publication \cite{owen23} claims to have found a ``design principle of 3D epigenetic memory systems". The proposed mechanism is essentially identical to the previous model but the performance of the model has been dramatically improved by assuming that the number of enzymes is smaller than the number of nucleosomes, compare 
Figs.~3B and E in Ref.~\cite{owen23}. 
The crucial element is that faithful restoration is contingent on the balance between missing marks and available methylases that deplete during the process, i.e., the mechanism requires fine-tuning.
Moreover, also this model assumes instantaneous methylation. The unphysical freezing of the chromosome after the collapse allows to ``transfer'' the epigenetic information in the sequence of methylated and non-methylated nucleosomes into a frozen chromosome conformation. However, as mentioned in requirement B, experiments suggest that recovery of the missing marks is an extremely slow process: reaching the original level of methylation after duplication-induced depletion takes about 20 hours \cite{xu12}. As the chromosomes are highly dynamic \cite{saintillan18}, it seems unlikely that their conformations can carry any such information even on a time scale of seconds. Furthermore, 
chromosome conformations are completely altered when chromosomes enter the mitotic state a few hours after DNA duplication: within minutes, the distinction between eu- and heterochromatin compartments is lost \cite{gibcus18}. This strongly suggests that the memory of the previous epigenetic state cannot be stored inside the chromosome conformation and that we do not understand yet why epigenetic information can be robustly transferred through 50 cell generations.

In the following, we put forward a new scheme for the restoration of epigenetic marks which does not suffer from these shortcomings, as all the information of the previous epigenetic state lies throughout the whole process in the diluted sequence of marks. Unlike in previous models \cite{sandholtz20,owen23}, where the chromosome conformational change is mainly driven by the attraction between the heterochromatic nucleosomes, in our model the polymer only ``assists'' in the formation of a self-assembled well-defined mesoscopic reaction vessel. This feature gives the necessary robustness to a scenario where a crucial system parameter, the fraction of methylated nucleosomes, changes through the process.

Specifically, we claim that the basic underlying physical mechanism is polymer-assisted condensation (PAC), a mechanism we discovered recently \cite{sommer22}. In this process, molecules, here HP1, with self-attraction 
are initially
in the mixed state, as their concentration is too low to spontaneously demix through liquid-liquid phase separation. 
Addition of a polymer with attraction to the molecules, here the heterochromatin stretches of a chromosome, can induce the spontaneous formation of a droplet that contains the attractive polymer sections. The droplet stops growing once it contains all the relevant polymer sections. The result is a micellar structure with heterochromatin inside the droplet and the euchromatin stretches forming external loops. 
% outside. 
PAC-induced condensates belong to the large class of macromolecular condensates which have been found in the cytoplasm and inside the nucleus of cells and which fulfill various biological functions \cite{banani17}. PAC is one of the mechanisms that can explain how the right droplets with the right composition and size occur at the right place in the cell.

Our hypothesis is supported by \textit{in-vitro} experiments demonstrating HP1 to exhibit liquid-liquid phase separation at sufficiently high densities~\cite{strom17,larson17}. This phase separation is induced by the attraction between non-structured regions in these proteins.  
In addition, as mentioned above, HP1 has a specific binding site to the H3K9me3 mark \cite{bannister01,nakayama01,lachner01}, suggesting that heterochromatin can play the role of the condensate-inducing polymer.

The goal of this study is to demonstrate that PAC of HP1 can indeed explain the propagation of epigenetic states of cells through 50 cell generations. To achieve this, we simulate a block copolymer which represents a model chromosome under various conditions of the parameters. We choose a rather small and simple system to be able to scan the parameter space in order to demonstrate the robustness of the system in producing reaction vessels through PAC. By setting up a simple methylation scheme with a reaction speed that depends on the local HP1 concentration, we demonstrate that heterochromatin marks lost
during duplication can indeed be reliably reconstructed. We propose that the well-defined droplet surface constitutes the physical manifestation of the boundary elements, originally postulated in Refs.~\cite{bannister01,gaszner06,barkess12}. Importantly, throughout the process, the droplet retains its liquid state and the polymer shows the expected dynamic fluctuations. Interrupting the process, as it 
occurs, for example, 
during chromosome compaction and separation, followed by the initiation of an uncorrelated  
polymer configuration does not hamper the restoration of the epigenetic marks.

%\clearpage

%%%%%%%%%%%%%%%%%%%%%%%%%%%%%%%%%%%%%%%%%%%%%%%%%%%%%%%%%%%%%%%%%%%%%%%%%%%%%
%%% PAC
%%%%%%%%%%%%%%%%%%%%%%%%%%%%%%%%%%%%%%%%%%%%%%%%%%%%%%%%%%%%%%%%%%%%%%%%%%%%%
\section*{Reaction container formation through PAC}\label{sec1}
PAC refers to the formation of a (protein) condensate due to the presence of a long flexible polymer (chromatin) in solution. In contrast to binding-induced interactions, where the protein forms temporary bridges between monomers, in PAC only a weak attraction due to the fluctuating polymer creates a chemical potential trap for the protein inside the polymer volume, causing a condensation transition. Thus, the principal driving force to form the condensate is the tendency of demixing of the protein in bulk. As a consequence, the precondition for PAC is a protein solution that is well above the critical point in the phase diagram but undersaturated, i.e., located outside of the miscibility gap, see Fig.~\ref{fig:PAC}(f). In practical terms, this means that the bulk solution displays spontaneous phase separation if sufficiently up-concentrated (\textit{in vitro}), a scenario common to virtually all proteins involved in biomolecular condensates and in particular observed for 
HP1, the scaffold component of the heterochromatin condensate \cite{strom17,larson17}.  In the following, we briefly recall essential results from the theory of PAC, for details the reader is referred to Ref.~\cite{sommer22}.

PAC can be understood and quantified considering a mean-field Flory-Huggins model \cite{degennes} with the free energy per volume unit and in units of $k_B T$:
\begin{eqnarray}
    F(c,\phi) &=& F_{el}(\phi) + F_{mix}(\phi, c) + \chi c(1-c-\phi) \nonumber \\ 
    &&+ \Pi -(\mu + \epsilon \phi)c\;.\label{eq:F_PAC_gen}
\end{eqnarray}
Here, $c$ and $\phi$ denote the volume fraction of the protein and the polymer respectively. The first two terms represent the chain elasticity and free energy of mixing of the protein and the polymer. Here, in particular, the Flory-Huggins form is applied as $F_{mix}=\frac{1}{n} c \ln c + (1-c-\phi)\ln (1-c-\phi)$, where $n$ denotes the volume of the protein with respect to the volume of the common solvent (water). The bulk interaction parameter $\chi$ has to be chosen above the critical value of the pure protein-water solution and it is useful to define $\eta=\chi-\chi_C>0$. The chemical potential of the bulk, $\mu$,  has to be set to be outside of the miscibility gap so that the protein is not phase separating by itself. This means that $\mu<\mu_{0}$, where $\mu_0$ denotes the value at coexistence. The equilibrium of the condensate with the bulk solution with respect to volume changes is given by the osmotic pressure, $\Pi$, of the proteins in the bulk. We note that $-\Pi$ represents the free energy per unit volume of the bulk at given chemical potential. $F(c,\phi)$ is thus actually the difference between the free energy density of the condensate and the bulk. The preference of the polymer with respect to the protein is modelled by $\epsilon$ in a mean-field-like manner. We note that if $\mu + \epsilon \phi>\mu_0$ condensation is favored due to the polymer-protein coupling. The equilibrium state is obtained by minimization of the free energy per monomer (since the number of monomers is a conserved quantity) with respect to $c$ and $\phi$. 

While the minimization problem $F(c,\phi)/\phi\to \text{min}$ can only be solved numerically in the general case, in Ref.~\cite{sommer22}  we developed an analytical solution for  $n=1$ using a Landau approximation of the free energy function.  This allows to calculate the PAC-transition in the form of
\begin{equation}\label{eq:pac_transition_LatticeGas}
\epsilon_{PAC}=\left[\frac{\lvert\mu\rvert}{\delta_{0}\left(1-\delta_{0}\right)}\right]^{1/2}~,
\end{equation}
with $\delta_0^2=\frac{3}{2}\eta$.  This defines a surface in the 3D parameter space ($\rho,\chi,\epsilon$), with $\mu=\ln\left(\frac{\rho}{1-\rho}\right)-\chi\left(2\rho-1\right)$ and $\mu_0 = 0$, where $\rho$ denotes the volume fraction of the protein in the bulk. Within this approximation, we also calculate the region of the PAC-state outside of the miscibility gap of the bulk solution, i.e. $\chi(\rho,\epsilon)$. The result is plotted for the case of $\epsilon=1$ in Fig.~\ref{fig:PAC}(f). For $n>1$ the essential difference is that then $\mu_0<0$ and $|\mu|$ has to be replaced by $|\Delta \mu| =| \mu-\mu_0 |$.

%\begin{multicols}{2}

A surprising result of the theory regards the volume of the condensate which is given by
\begin{equation}\label{eq:R3}
    V_{PAC}\simeq N\frac{\lvert\mu\rvert^{-1/2}}{\left(\delta_{0}\left(1-\delta_{0}\right)\right)^{1/2}}~,
\end{equation}
with a prefactor of order unity. What is particularly noteworthy is that the condensate volume is independent of the strength of the monomer-protein interaction, $\epsilon$. This is due to the fact, that the droplet is dominated by the condensed protein which saturates the polymer, i.e., the monomers are surrounded by proteins in the PAC state. This means that a stronger protein-monomer interaction does not have much impact on the net interaction inside the droplet once the condensate is formed. Indeed this prediction is confirmed in the simulations \cite{sommer22}. This aspect is important in the context of epigenetic restoration because after replication only half of the epigenetic markers are left on a given daughter chromosome and therefore the effective value of $\epsilon$ is now $\epsilon/2$. But as long as $\epsilon/2>\epsilon_{PAC}$ we still have the container for the necessary enzymatic reactions and, to a good approximation, this container maintains its volume throughout the whole re-methylation process.

\begin{figure}[ht]
     \centering
    \includegraphics[width=18.0cm]{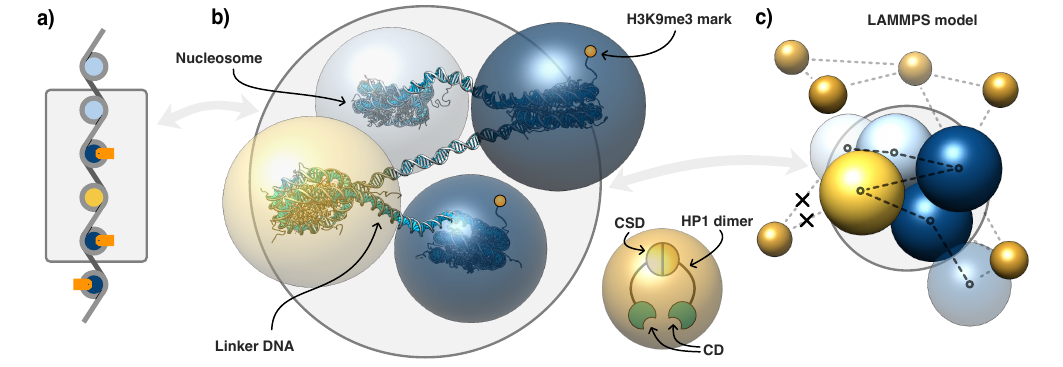}
    %\caption{\textbf{Coarse-grained representation of chromatin:} (a) Each nucleosome (147 bp of DNA wrapped around a histone octamer) is modelled as a hard sphere. Linker DNA (here of length 50 bp) is treated as the bond between two of these hard spheres. Two nucleosomes feature methylated marks on histone tails. Spheres are coloured according to their identities: dark blue denotes methylated heterochromatin nucleosomes, yellow heterochromatin nucleosomes missing marks, and light blue euchromatin nucleosomes. (b) HP1 dimer modelled as a free bead. HP1 molecules bind together via chromo shadow domains (CSD) to form a dimer. Chromo domains (CD) bind to methylated nucleosomes.
    %\esc{@Sukanta Remember to change the caption.}
    %}
    \caption{\textbf{Coarse-grained representation of chromatin and HP1s:} (a) Schematic representation of chromatin section in the same style as in Fig.~\ref{fig:general_scheme}. (b) Representation of the same section as in (a) in the computer model. Each nucleosome (147 bp of DNA wrapped around a histone octamer) is modelled as a hard sphere. Linker DNA (here of length 50 bp) is treated as the bond between two of these hard spheres. Two nucleosomes feature methylated marks on histone tails. Spheres are coloured as follows: dark blue denotes methylated heterochromatin nucleosomes, yellow heterochromatin nucleosomes missing marks, and light blue euchromatin nucleosomes. HP1 dimers are represented by free beads (orange). HP1 molecules bind together via chromo shadow domains (CSD) to form a dimer. Chromodomains (CD) bind to methylated nucleosomes. (c) Section of chromatin surrounded by HP1 molecules as treated in LAMMPS molecular dynamics simulation. The nucleosome beads have bonded interactions (FENE) in between, which are shown by black dashed lines. HP1 beads have attractive interactions with themselves and with dark blue nucleosomes (grey dashed lines).}
    \label{fig:CG} 
\end{figure}

Here we carry out molecular dynamics simulations to study the formation of an HP1 condensate through PAC with a block copolymer representing a model chromosome (see Methods for details). In short, the chromosome is represented by a bead-spring model, and HP1 molecules are modeled by unconnected beads (of the same diameter as the monomers). The coarse-grained representations of chromatin and of the HP1 dimer are shown in Fig.~\ref{fig:CG}(b). The interactions in our model involve attractive LJ potentials between the HP1-beads of strength $\chi_{S}$, and between the heterochromatically marked monomers and HP1 of strength $\epsilon_{S}$, see Fig.~\ref{fig:CG}(c); the index “S” indicates the values used in the simulation model. Interactions between monomers (regardless of whether they are marked or not) and between unmarked monomers and HP1 are modeled by purely repulsive potentials. All lengths are measured in units of the bead diameter. Throughout our simulations we use $\chi_{S}=1.1$, a value well above the critical point $\chi_{X}\simeq0.9$ \cite{watanabe12}. In Ref.~\cite{sommer22} we have simulated PAC for a homopolymer with a chain length of 300 monomers. The phase diagram determined in that study is shown in Fig.~\ref{fig:PAC}(e), displaying the plane $\epsilon_{S}$ versus $c_b$ (the bulk volume fraction of HP1) for fixed $\chi_{S}=1.1$. The red vertical line indicates the transition to the condensed phase which occurs at volume fraction $c_{b}\simeq0.0575$.

\begin{figure}[h!]
    \centering
    \includegraphics[width=18.0cm]{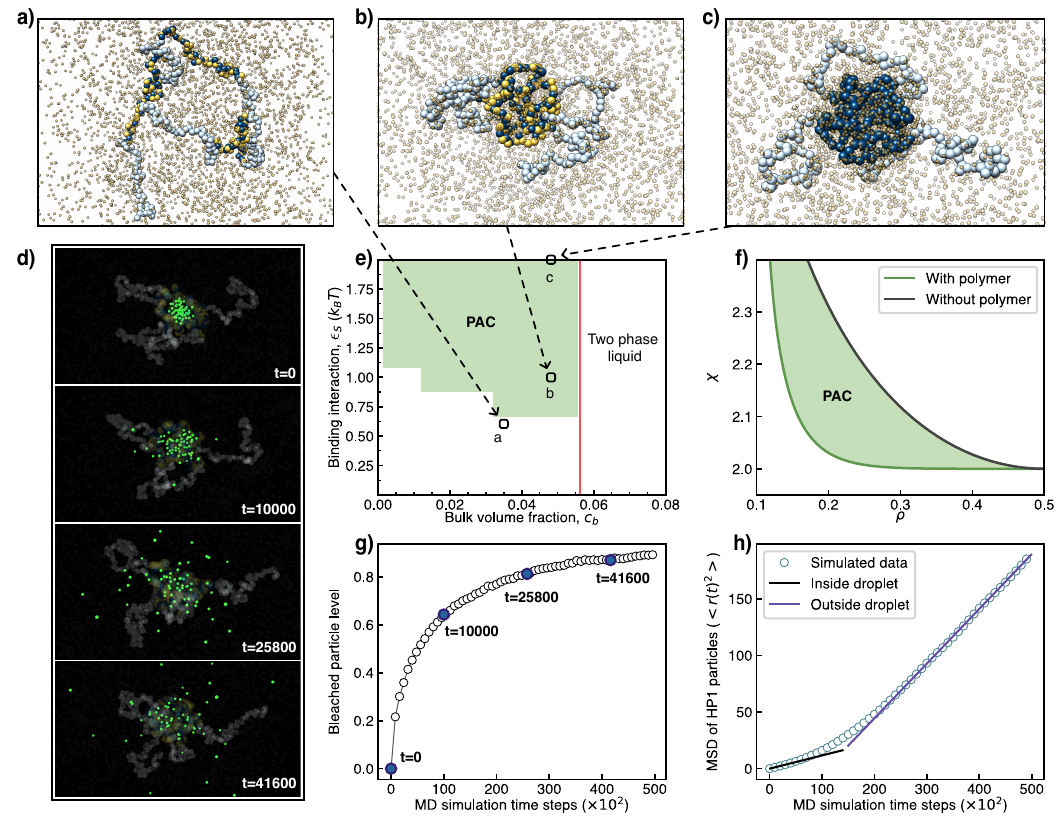}
    \caption{\textbf{Model chromosomes induce liquid condensates:} (a)-(c) Snapshots of block copolymers containing marked (dark blue) and unmarked (light blue) nucleosomes. (b) and (c) show cases that lead to condensates through PAC. In both cases $c_b=0.048$ and $\epsilon_{S}=2$ but for (b) one has effectively $\epsilon_{S}=1$, as every second monomer has lost its epigenetic mark (yellow monomers). (a) shows a system with a lower HP1 bulk concentration, $c_b=035$, and with $\epsilon_{S}=1.3$ but with half of its markers missing and thus effectively $\epsilon_{S}=0.65$. This system is outside the PAC region and indeed no condensate is formed. (d) Snapshots of configurations of HP1s, at different points in time. Highlighted in green are the HPs that formed the droplet at $t=0$. (e) Phase diagram determined from molecular dynamics simulation for $\chi_{S}=1.1$ \cite{sommer22}. (f) Phase diagram of a two-component solution (e.g. HP1 and water) showing a miscibility gap to the right of the black curve. This gap is widened to the left (beige region) in the presence of a polymer. Calculated from Eq.~\ref{eq:pac_transition_LatticeGas} for the case of $\epsilon=1$.  (g) Fraction of HPs from the bulk (at $t=0$) inside the droplet as a function of time. (h) MSD of the HP1s which started inside the droplet at $t=0$.
    }
    \label{fig:PAC}
\end{figure}

In the current study, we use a model chromosome consisting of seven blocks. Four blocks, including the outer blocks, represent euchromatin, and three blocks are heterochromatin. Each block is 50 monomers long which corresponds to the median length of blocks with H3K9me3 nucleosomes in humans \cite{barkess12}. We thus interpret 
each monomer of the polymer as one nucleosome, see Fig.~\ref{fig:CG}(b).
In Figs.~\ref{fig:PAC}(a)-(c) we show three example conformations of the model chromosome for three different parameter sets, indicated by the arrows pointing at the corresponding positions in the phase diagram, Fig.~\ref{fig:PAC}(e). The system in (c) features fully methylated blocks shown in dark blue and non-methylated blocks in light blue. The parameters are chosen such that the system lies deep inside the PAC region, see arrow to Fig.~\ref{fig:PAC}(e). We observe that the heterochromatin stretches induce a condensate which is surrounded by a corona of euchromatin stretches. In case (b), we have removed half of the marks from the heterochromatin blocks. The corresponding monomers are highlighted in yellow but are identical with respect to their interactions with the light blue monomers. As half of the marks are missing, the interaction strength between HP1 and the heterochromatin blocks is effectively only half as strong. The arrow to the phase diagram takes this into account and shows that the system is still inside the PAC region. As expected, we find again a micellar structure with a central condensate. Snapshot (a) shows an example where the system is outside the PAC region. We note that similar micellar structures have been put forward in Ref.~\cite{adame23} but here the focus was on changes in outer loops due to chromatin-binding proteins such as RNA polymerase II.

We next study the dynamics of the condensate. In Figs.~\ref{fig:PAC}(d), (g), and (h) we track the HP1 molecules over time. Specifically, in the snapshots of Fig.~\ref{fig:PAC}(d) we highlight HP1s which are part of the droplet at $t=0$ in green and then show their positions after different amounts of time steps. This approach shares some similarities to FRAP (fluorescence recovery after photobleaching) experiments. As can be seen from the snapshots, there is a fast exchange of HP1s
with the bulk so that the outline of the droplet cannot be recognized anymore for the two larger times. This can also be read off the curve in Fig.~\ref{fig:PAC}(g) which shows the fraction of the ``bleached'' HP1s
inside the droplet, i.e., the 
HP1s
that are not part of the droplet at $t=0$. Already at $t=25800$ about $80\%$ of the droplet material is composed of these bleached HP1s. 
To further quantify the dynamics,
we show in Fig.~\ref{fig:PAC}(h) the mean-squared displacement (MSD) of the marked beads as a function of time. Around $t=15000$ the data show a crossover from a slow diffusion regime with a diffusion constant of $D\approx 0.12$ to a fast diffusion regime with $D\approx 0.51$. This crossover reflects the escape of the marked HPs from the droplet to the bulk. It is important to note that the HP1 dynamics inside the droplet is quite high since the diffusion constant is only about four times smaller than in the bulk. Moreover, also the polymer is highly dynamic, including the heterochromatin sections inside the condensate which remain mobile showing Rouse dynamics throughout the entire cycle of mark-restoration, see Methods for details.

Overall, a dynamic picture emerges of a condensate that rapidly exchanges proteins with the surroundings and that contains highly fluctuating polymer sections. This dynamics creates opportunities and challenges. On one hand, it allows other factors like methylases to access the heterochromatin sections. On the other hand, the system cannot rely on a frozen polymer configuration to remember and reconstruct the previous epigenetic state. Rather, at all times epigenetic memory is stored exclusively in the sequence of labeled and unlabeled nucleosomes. Results from computer simulations presented in the next sections demonstrate that, despite the periodic dilution of marks, this highly dynamic system can remember its original epigenetic state even after 50 cell generations.

%\clearpage
%\section*{Setting up the methylation reaction scheme}\label{sec2}
%\input{Sections/Setting up the methylation reaction scheme}
%\clearpage

%%%%%%%%%%%%%%%%%%%%%%%%%%%%%%%%%%%%%%%%%%%%%%%%%%%%%%%%%%%%%%%%%%%%%%%%%%%%%
%%% SINGLE CYCLE
%%%%%%%%%%%%%%%%%%%%%%%%%%%%%%%%%%%%%%%%%%%%%%%%%%%%%%%%%%%%%%%%%%%%%%%%%%%%%
\section*{Restoration of epigenetic marks in one cell generation}\label{sec3}

In this section, we demonstrate that our system allows for the restoration of missing epigenetic marks 
within a single cell generation  
and in the next section, we extend this analysis to 50 cell generations.
These simulations are meant as a proof of concept, performed for one rather arbitrarily chosen set of parameters. Later in this paper, we show that this scenario is robust in the sense that it works over a large range of parameters, which supports the plausibility of our approach.

\begin{figure}
    \centering
    \includegraphics[width=18cm]{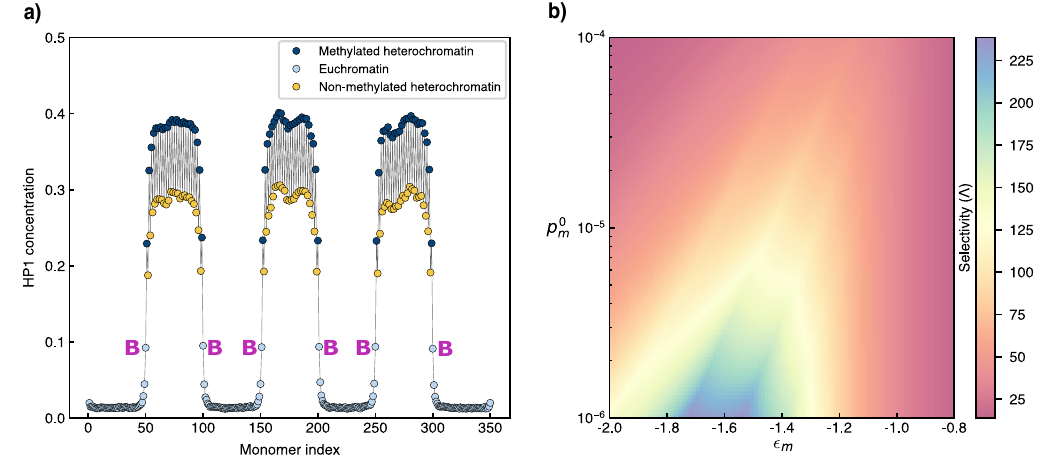}
    \caption{\textbf{HP1 concentration and methylation probability:} (a) Local HP1 concentration profile around monomers of model chromosome. Heterochromatin has alternating patterns of retained (dark blue circles) and missing marks (yellow circles). The number concentration is calculated in a sphere of radius $r_{\text{c}}$=1.5 around each monomer, averaged over 2000 equilibrium ensembles. Euchromatic nucleosomes next to heterochromatin domains are labelled by the letter ``B''. (b) Map of selectivity $\Lambda$ in the $p_m^0$--${\epsilon}_m$ parameter space (see text for details).}
    \label{fig:selectivity}
\end{figure}

Our starting point is a chromosome micelle as described above with the parameters $\chi_{S}=1.1$, $\epsilon_{S}=2$ and $c_b=0.048$, corresponding to the example shown in Fig.~\ref{fig:PAC}(b).
With a chromosome micelle containing missing epigenetic marks in place, methylation reactions must be performed to restore these marks. For this mechanism to be effective, the reactions for nucleosomes lacking epigenetic marks within heterochromatin blocks must occur significantly faster than those in euchromatic blocks. The key challenge in our model arises from the fact that both types of unmarked nucleosomes are identical. Therefore, the distinction can only lie in differences of the local HP1 environment, as illustrated in Fig.~\ref{fig:selectivity}(a).

This plot shows the number density $\rho_\mathrm{HP}$ of HP1 molecules 
in the vicinity of given monomers, counting all HP1 within a radius of $1.5\sigma$ from these monomers.
For simplicity, we choose  
a sequence of epigenetic marks in which the heterochromatin blocks consist of alternating patterns of marked nucleosomes (dark blue) and unmarked nucleosomes (yellow). 
The concentration profile exhibits sharp transitions that reflect the block-like organization of the model chromosome, with very low HP1 concentrations in euchromatin blocks and high concentrations in heterochromatin blocks. 
In addition, the alternating sequence of marked and unmarked nucleosomes within the heterochromatin blocks leads to periodic undulations in the HP1 concentration. Importantly, the HP1 concentrations of all yellow nucleosomes are higher than those of all euchromatic nucleosomes.

Note that differences in local HP1 concentrations between 
marked (dark blue) and unmarked
(light blue or yellow) nucleosomes are smallest at the boundaries between domains. Here relative differences in
concentrations are only of the order two. 
This indicates the need for a sharp dependence of methylation rates on the local HP1 environment. Here, we assume an exponential relationship between the rate and the local HP1 concentration. Consequently, the probability of a randomly selected nucleosome (if unmethylated) undergoing methylation is expressed as:
\begin{equation}\label{eq:meth_prb}
 p_m=p_m^0\exp(- n {\epsilon}_m),
\end{equation}
where $p_m^0$ is the base methylation probability of a nucleosome, related to the bulk concentration of enzymes,
${\epsilon}_m$ is a dimensionless parameter and $n$ is the number of HP1 dimers within a radius of $1.5\sigma$.
The functional dependence of $p_m$ is not crucial in our model, as long as it is steep enough and it should be regarded as a threshold or activation function. The specific form of Eq.~\ref{eq:meth_prb} might be interpreted as a preference of the methylase to be inside the HP1 condensate, which can be expressed as an effective attractive potential between HP1s and methylases.
% \smc{One additional fact might be relevant to mention here that the biochemistry of the methylase enzyme requires it to be `recognised' by the epigenetic marks for allosteric activation. This is well established in the following reference-} ~\cite{Muller16} \smc{An unactivated enzyme molecule has significantly smaller intrisic catalyzation rate than an activated one. Therefore the concentrated presence of enzymes around the heterochromatin sections may cause a faster, prior chemical reactions.}
Moreover, various studies have highlighted the propensity of the condensate environment to promote enzymatic activity~\cite{testa21,Muller16}, which further enhances the selectivity of reactions to take place inside the reaction container. In our model, these effects are collectively accounted for by the effective potential $\epsilon_m$.
% \smc{Furthermore, this proposition can be reinforced by the fact that the entire cell generation is very long (we are already saying this in the introduction) compare to the timescale of an enzymatic reaction; therefore it could be viable to assume the system to be in a quasi-equilibrium state. Hence formalism of equilibrium statistical mechanics still holds.}
% In addition, the reaction rate of an enzyme might increase due to the condensate environment, as e.g., observed in Ref.~\cite{testa21}. 
We note that since $p_m$ is a probability, its value is restricted to $p_m \leq 1$.
% \esc{This could already be included in the equation, e.g., as 
% \begin{equation}
%      p_m = \mathrm{min} \left\{  p_m^0\exp(- n {\epsilon}_m) , 1 \right\}
% \end{equation}}
% Whenever $p_m > 1$, we set $p_m=1$.

The methylation reaction rate, Eq.~\ref{eq:meth_prb}, adds two new parameters to our model, $p_m^0$ and $\epsilon_m$. These parameters need to be chosen such that the methylation rates of the yellow monomers in Fig.~\ref{fig:selectivity}(a) are large enough compared to the methylation rates of the light blue monomers at the boundary to heterochromatin (labelled by the letter ``B''), leading to a clear separation of time scales for the two types of monomers. This way the heterochromatin blocks are restored before the blocks start to grow into the euchromatic regions. To find promising values for $p_m^0$ and $\epsilon_m$, we introduce the selectivity $\Lambda$, the ratio of the average methylation probability of the yellow nucleosomes to the average methylation probability of the light blue nucleosomes at the boundaries. Details about $\Lambda$, including a more general definition which includes also other patterns of missing marks, are provided in Methods. In Fig.~\ref{fig:selectivity}(b), $\Lambda$ is plotted in the $p_m^0-{\epsilon}_m$ parameter space. We observe a broad range of parameters where the system is highly selective with values of $\Lambda$ in the range of 20 to 200. In the following we choose ${\epsilon}_m=-1.4$ and  $p_m^0=10^{-5}$.

Finally, we introduce one more parameter, the distribution bias $p_b$, which affects the pattern of missing marks at the beginning of the simulation. It is known experimentally that nucleosomes are distributed in a globally symmetric fashion  between the daughter DNA molecules, i.e., both daughter cells inherit half of the epigenetically marked histones from the parent \cite{yu18}, although the process is intrinsically asymmetric, as there is a leading and a lagging strand. Molecular details of the transfer 
remain elusive, but there is some evidence suggesting histones to be ``toggled'' between leading and lagging strands in a fairly regular manner 
facilitated by the fork protection complex \cite{charlton24}. In the current study, we allow for the whole range of scenarios of the local distribution, from strictly alternating to completely random.
We distribute the nucleosomes between the two DNA copies, refill the missing nucleosomes with unmarked nucleosomes, and then use only one of the two resulting sequences as an input for our simulation. 
Specifically, to create a diluted epigenetic sequence, we transfer the first nucleosome on the left of the entire chromsome chain to either of two newly formed strands with probability 1/2. This results into either keeping the mark or remove it (meaning the nucleosome is transferred to the other DNA copy).
We then assume the probabilities for the next nucleosome to end up on the same and other DNA copy to be $1/2-p_b$ and $1/2+p_b$, respectively.
The range of possible $p_b$-values goes from $p_b=0$, the completely random case, to $p_b=1/2$, the fully alternating pattern shown in Fig.~\ref{fig:selectivity}(a).
The more random the distribution, the larger is the probability of larger segments of nucleosomes with missing epigenetic marks (defects) which makes it more challenging to re-establish the marks. For our proof-of-concept study, we choose $p_b=0.4$, close to the alternating case but with some defects spanning over a length of two or more nucleosomes.
An analysis of the effect of $p_b$ on the re-establishment of marks is provided in the section on robustness.

\begin{figure}[ht]
     %\centering
    \includegraphics[width=18cm]{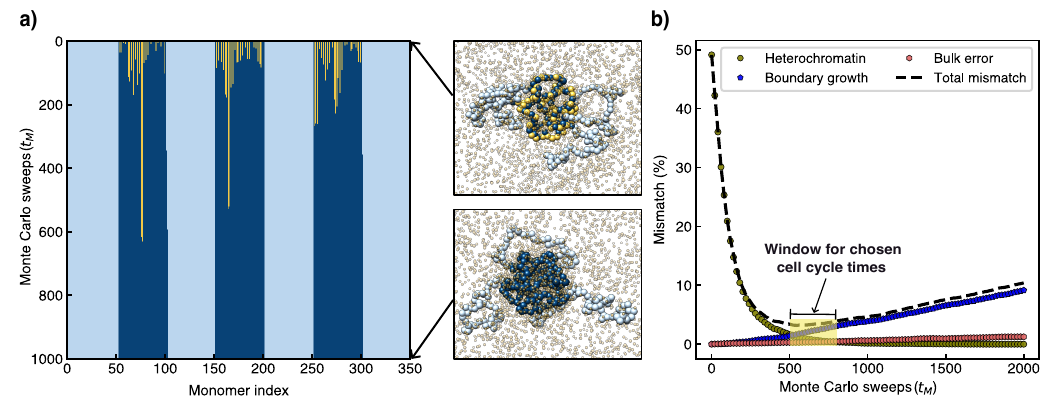}
    \caption{\textbf{Re-establishment of epigenetic marks in one cell generation:} (a) Methylation sequence of model chromosome as a function of MC cycle number. The initial sequence was generated with a distribution bias of $p_b=0.4$. The parameters for the methylation reaction are $\epsilon_m=-1.4$ and $p_m^0=10^{-5}$. Snapshots show the system at the beginning and end of the re-establishment run. (b) Mismatches between the current and target sequence as a function of the number of MC cycles. The black dashed line shows the non-monotonic behavior of the total mismatch with a minimum at around $t_{M}=540$. The profiles are averaged over 100 independent simulations with different initial configurations. The shaded region in yellow indicates a suitable window for choosing the cell cycle time.
    }
    \label{fig:one gen} 
\end{figure}

We are now in the position to study the methylation dynamics for a single cell generation. First, the model chromosome, initially containing heterochromatin blocks with yellow defects of missing marks, is equilibrated in an MD run. Then starts a series of Monte Carlo (MC) sweeps, each containing two steps: a set of methylation attempts followed by an MD equilibration run, see methods for details. This way the droplet-polymer system 
is given time to relax
its spatial conformation to the changing epigenetic sequence. 

Figure \ref{fig:one gen}(a) displays the evolution of the epigenetic sequence from top to bottom. We observe that with increasing time more and more yellow nucleosomes get methylated, with most marks being re-established after as few as 200 MC sweeps, $t_{M}=200$. The last two yellow nucleosome stretches, both two nucleosomes long, are methylated shortly after $t_{M}=500$ and 600. Importantly, within the same time range, there is hardly any growth at the boundaries of the heterochromatin domains, demonstrating that our system features the required separation of time scales between methylating right and wrong nucleosomes.

This separation of time scales is further illustrated in Fig.~\ref{fig:one gen}(b), where we show the percentage of mismatch between the parent epigenetic sequence and that of the diluted sequence as a function of the MC sweeps, averaged over 100 independent simulations, see Methods for details. The mismatch of the heterochromatin starts at $50\%$ and decays rapidly with the MC sweeps. Moreover, there is a slowly linearly growing mismatch in the euchromatic fraction of the chromosome which has two contributions: the dominant boundary growth (the growth of heterochromatin into the euchromatin) and a very slow bulk error, stemming from heterochromatin that spontaneously forms within the euchromatin sections. 
The minimum in the total number of mismatches is reached around $t_{M}=550$.

%\clearpage

%%%%%%%%%%%%%%%%%%%%%%%%%%%%%%%%%%%%%%%%%%%%%%%%%%%%%%%%%%%%%%%%%%%%%%%%%%%%%
%%% MULTI CYCLE
%%%%%%%%%%%%%%%%%%%%%%%%%%%%%%%%%%%%%%%%%%%%%%%%%%%%%%%%%%%%%%%%%%%%%%%%%%%%%
\section*{Epigenetic marks restoration through many cell generations}\label{sec4}
We now extend our proof-of-concept simulation to 50 cell generations, i.e., into the range of the Hayflick limit \cite{hayflick65}. To do so, we repeat the simulation from the previous section 50 times. At each generation, we stop the simulation after $t_{C} = 600$ MC sweeps, approximately when the mismatch reaches its minimum in Fig.~\ref{fig:one gen}(b). We pick the epigenetic sequence from that last sweep as input for the next cell generation. Specifically, we remove half of the marks with a distribution bias $p_b=0.4$ from that sequence, restart the system with a random configuration, perform an equilibration run during which a new condensate forms, and then run again for $t_{C}$ MC sweeps.

\begin{figure*}[ht]
     %\centering
    \includegraphics[width=18cm]{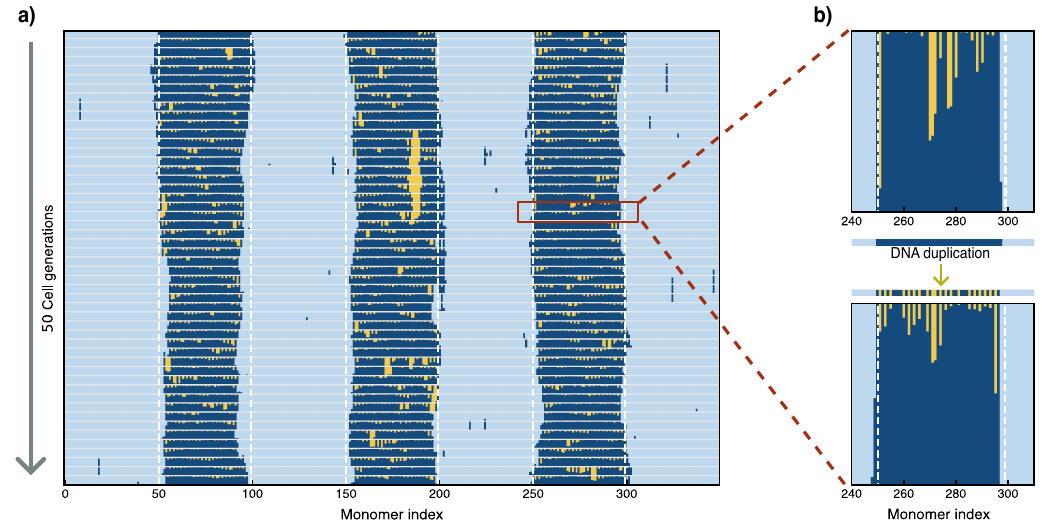}
    \caption{\textbf{Multi-cell generation simulations of epigenetic restoration:} 
    (a) Epigenetic mark dilution and re-establishment through 50 cell generations. The original positions of the six euchromatin/heterochromatin boundaries are marked with dashed white vertical lines. The restoration profile shows fairly stable heterochromatin domains with some bulk error arising rarely. (b) A zoomed-in view of two consecutive cell generations. Between two cell generations, the sequence from the end of the previous generation is diluted by removing half of the marks (mimicking the effect of DNA duplication). The resulting sequence serves as input for the next cell generation.
    }
    \label{fig:mcg1} 
\end{figure*}

The resulting time development of the epigenetic sequence is shown in Fig.~\ref{fig:mcg1}. We observe that even after 50 generations, the block-like arrangement of eu- and heterochromatin is still intact, including the sizes and positions of the domains. This is remarkable since each time when half of the marks are removed, a given heterochromatic domain shrinks by an average of one nucleosome. One might thus expect that the domains, at the start each being 50 nucleosomes long, disappear after 50 cell generations. However, Fig.~\ref{fig:mcg1} lets us conclude that the shrinkage of the domains is roughly cancelled by an outward drift of heterochromatin into the euchromatic domains. As a result, the boundaries show diffusive trajectories, as can be seen by comparing the boundary positions over time with the original boundaries, indicated by white dashed lines. Our choice of the number of MC sweeps per generation, chosen to minimize the total error per generation, might play a role here.

Another noteworthy observation in Fig.~\ref{fig:mcg1} is the emergence of a long-lived defect, characterized by a wide gap of unmethylated heterochromatin. This defect first appears in the middle block in the 12th cell cycle generation as a sequence of three unmarked nucleosomes. It persists over several subsequent generations, eventually expanding to a size of six missing marks by the 17th generation. 
However, during the 21st generation, the defect rapidly shrinks to a single unmarked nucleosome and disappears entirely by the 22nd generation. This observation suggests that the re-establishment mechanism is capable of ``healing" larger defects, though the process may require several cell generations, as larger defects have the tendency to reside at the surface of the droplet.

We occasionally observe the spontaneous formation of isolated marks within euchromatin domains. The vast majority of these defects remain isolated and are never found to exceed two marks. Notably, all occurrences are found to disappear after a few cell generations. This is a consequence of cell division-induced mark dilution, which stabilizes euchromatin domains even in the absence of enzyme-mediated catalytic mark removal.
%\clearpage

%%%%%%%%%%%%%%%%%%%%%%%%%%%%%%%%%%%%%%%%%%%%%%%%%%%%%%%%%%%%%%%%%%%%%%%%%%%%%
%%% ROBUSTNESS
%%%%%%%%%%%%%%%%%%%%%%%%%%%%%%%%%%%%%%%%%%%%%%%%%%%%%%%%%%%%%%%%%%%%%%%%%%%%%
\section*{Robustness of restoration scenario}\label{sec5}
So far, we have provided a proof of concept for the re-establishment scenario for one set of parameters. In the current section, we give evidence that the scenario is rather robust with respect to changes in the parameters. We start by reiterating the point that PAC itself is already very robust. As can be seen in Fig.~\ref{fig:PAC}(e) and (f), there is a large range of parameters where PAC occurs. Moreover, as the free energy of the polymer-condensate is dominated by the condensate proteins, many quantities, 
such as the condensate volume, Eq.~\ref{eq:R3}, do not depend on ${\epsilon}_S$, the attraction strength between the marked monomers and HP1. In fact, the map of selectivity $\Lambda$ in the $p_m^0$--${\epsilon}_m$ parameter space in Fig.~\ref{fig:selectivity} (for ${\epsilon}_S=2$)
shows little change as we vary this value from ${\epsilon}_S=1.7$ to 2.5 in the Extended Data Fig.~\ref{fig:selectivity_variation_eps_S}. 
The selectivity only drops dramatically once the system approaches the boundary between PAC and the mixed state, which is the case for ${\epsilon}_S=1.5$.
Thus the selectivity is closely linked to the robustness of PAC.

We also note that the re-establishment process can be interrupted at any moment (e.g., by
switching off the attraction between HP1 and the methylated nucleosomes) and then be restarted (e.g., by 
switching the attraction on again) without compromising the faithful re-establishment of the epigenetic sequence. An example is provided in the Extended Data Fig.~\ref{fig:cell_cycle_stopped}. This feature is important since the re-establishment of the markers must be interrupted during the dramatic rearrangement of chromosomes during mitosis, in which the distinction between eu- and heterochromatin is temporarily disrupted \cite{gibcus18}.

\begin{figure*}[ht]
     \centering
    \includegraphics[width=8.8cm]{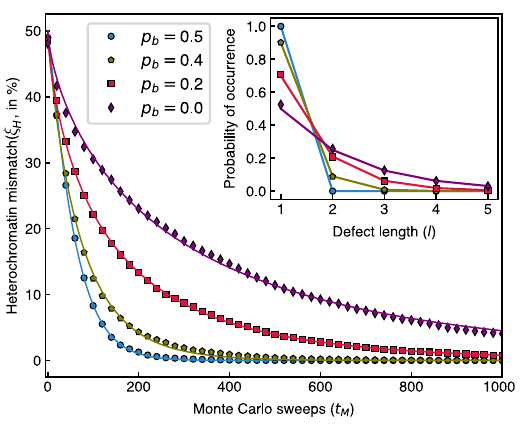}
    \caption{\textbf{Decay of heterochromatin mismatch with time for different values of $p_b$:} The plot shows that the re-establishment speed varies strongly with $p_b$. Each  plot is obtained by averaging over 100 simulations. Solid lines were obtained by fitting Eq.~\ref{eq:mismatch} to the data. Inset: Probability distributions of defect sizes for different values of $p_b$. The data points are obtained from simulating the dilution process on the chromosome model whereas the solid lines represent the theoretical estimate for the probability $(1/2-p_b)^{l-1}(1/2+p_b)$ for a defect of length $l$ to occur (at the start of a defect, the defect continues for exactly $l-1$ additional nucleosomes with probability $(1/2-p_b)^{l-1}(1/2+p_b)$); this estimate works well for defect lengths $l$ that are much shorter than the length of a heterochromatin domain.
    }
    \label{fig:stretched_exp} 
\end{figure*}

We focus next on the robustness of the scenario on parameters related to mark dilution and reaction speed. In the proof-of-concept simulation, we assumed that nucleosomes are distributed between the two DNA copies with a distribution bias of $p_b=0.4$. 
The bias affects the distribution of defects inside the heterochromatin, as shown in the inset of Fig.~\ref{fig:stretched_exp} 
which depicts the probability of occurrence of different-sized defects for various values of $p_b$.
While $p_b=0.5$ produces an alternating pattern where all defects consist of a single nucleosome, reducing $p_b$ to 0.4 results in defects of length two in approximately $10\%$ of cases.

As suggested before by Fig.~\ref{fig:mcg1}, defects spanning multiple marks may require several cell generations to be repaired.
This can be clearly seen when inspecting the selectivity for chromosomes with alternating blocks of marked and unmarked nucleosomes of different lengths. 
While single-block defects exhibit high selectivity (see Fig.~\ref{fig:selectivity}), larger blocks show significantly reduced selectivity, as shown in Extended Data Fig.~\ref{fig:selectivity_variation}(d) and (e).
This raises the question of whether, in scenarios with greater randomness than $p_b=0.4$, our model can still successfully re-methylate heterochromatin before epigenetic marks encroach into the euchromatin domains. To address this, we examine the evolution of the heterochromatin mismatch as a function of the MC cycle, shown in Fig.~\ref{fig:stretched_exp}. As evident from the plot, the decay of the heterochromatin mismatch slows significantly with decreasing distribution bias $p_b$. Notably, these curves deviate from simple exponential behavior but are well approximated by compressed or stretched exponentials:

\begin{equation}\label{eq:mismatch}
    {\xi}_H={{\xi}_H}^0 \exp \left( -({t_M}/{\tau})^{\nu} \right)
\end{equation}
\noindent
where the exponent $\nu$ depends on $p_b$; ${{\xi}_H}^0=50$ in all the cases. 
In the alternating case ($p_b=0.5$), the exponent $\nu$ is slightly greater than one ($\nu \approx 1.12$), suggesting the droplet to become denser during the mark re-establishment process, a second-order effect in the PAC scenario. For all other simulated cases, the exponent is less than one, with $\nu=0.89$ for $p_b=0.4$, $\nu=0.71$ for $p_b=0.2$, and $\nu=0.69$ for $p_b=0.0$. This reflects the fact that epigenetic sequences of heterogeneous defect sizes are characterized by locally different methylation probabilities, i.e., larger defects require more time for re-methylation.

\begin{figure*}[!ht]
    \centering
    \includegraphics[width=17.5cm]{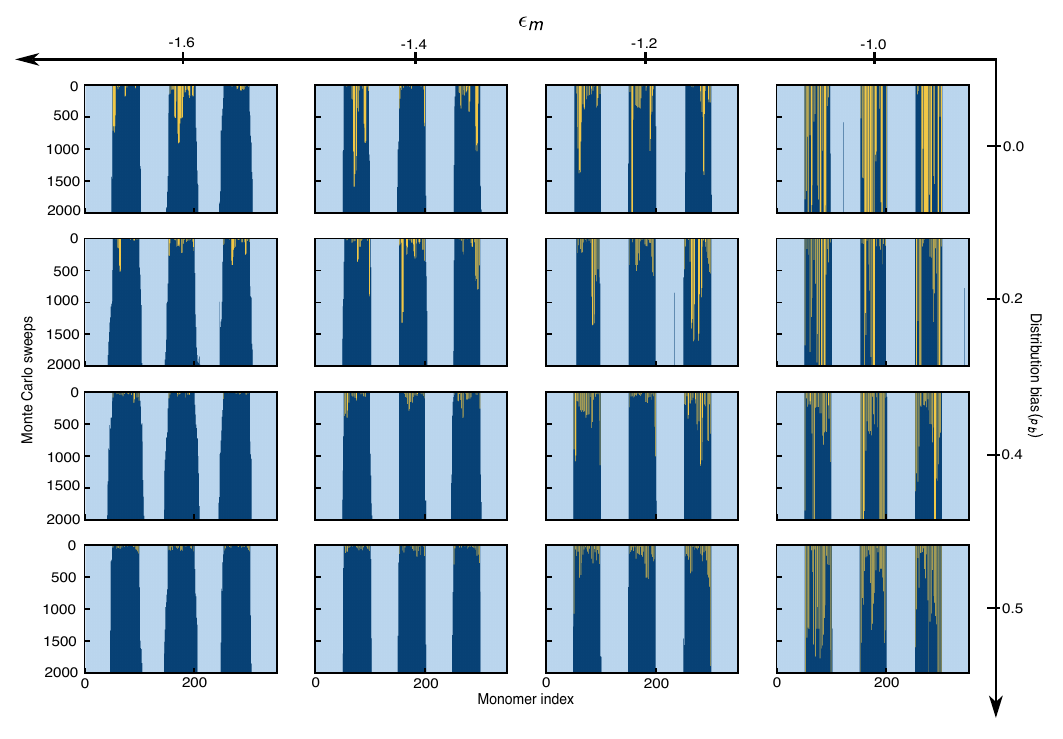}
    \caption{
    \textbf{Single cell cycle profiles across a vast parameter landscape:}
    Evolution of methylation status of nucleosomes through a cell cycle for different values of ${\epsilon}_m$ from Eq.~\ref{eq:meth_prb} and of the distribution bias, $p_b$. As can be seen from the various plots, the restoration mechanism works in the whole range of parameters displayed, yet the time scales for restoration differ strongly.
    }
    \label{fig:enter-label}
\end{figure*}

Next, we test whether our scenario is still able to repair the defects before heterochromatin spreads into euchromatin, despite the much longer recovery timescales for more random sequences. To do this, we study in Fig.~\ref{fig:enter-label} the evolution of epigenetic sequences for a single cell generation for a long period of 2000 MC sweeps for the $p_b$-values discussed above. In addition, we also vary the value of ${\epsilon}_m$ from Eq.~\ref{eq:meth_prb}, namely ${\epsilon}_m=-1.0$, $-1.2$, $-1.4$ (the previously used value), and $-1.6$. As can be seen from the various plots, the defect closing occurs faster for larger $p_b$-values, i.e., smaller defect sizes, and more negative ${\epsilon}_m$-values. This reflects the fact that nucleosomes inside smaller defects will be surrounded by more HP1 molecules and methylation rates increase exponentially with ${\epsilon}_m$. Notably, however, within the whole range of parameter values considered in Fig.~\ref{fig:enter-label}, the time scale for closing the defects is faster than the time scale for growing into euchromatin, suggesting that the scenario works robustly everywhere. 

Finally, we discuss the robustness of the scenario with respect to the cell cycle time. In the Extended Data Fig.~\ref{fig:many} we show various examples of time developments of epigenetic sequences for 50 cell generations. In all examples, we use again the parameters from our proof-of-concept study 
with the exception of
the cell cycle time (counted by the number of MC cycles). Specifically, we 
consider
$t_{C}=500$, 550, 580, 600 (the value used in Fig.~\ref{fig:mcg1}), 700 and 800. For the shortest cell cycle time we observe a slight inward drift of the heterochromatin boundaries and for the longest two an outward drift. Nevertheless, for the whole range of values considered here, the starting sequence can still be recognized after 50 cell generations. 

It is important to mention that we were only able to perform a small number of 50-generation runs due to the long simulation times. This means that we cannot comment on the success rate of our mechanism to re-establish the starting sequence due to lack of statistics. In Extended Data Fig.~\ref{fig:error}, we show two examples where the epigenetic sequence changed substantially. Subplot (a) depicts a simulation run with a cell cycle time of 500 where the leftmost heterochromatin domain almost disappeared. In subplot (b), a run with a cell cycle time of 800, a new domain formed early between the middle and rightmost domains, grew and eventually connected the two domains. This indicates two missing elements in our scenario: one that makes the re-establishment of epigenetic marks independent of cell cycle time, and one that suppresses the spontaneous formation of new heterochromatin domains. 
We further discuss the possible nature of these missing elements at the end of the discussion section.

%\clearpage

%%%%%%%%%%%%%%%%%%%%%%%%%%%%%%%%%%%%%%%%%%%%%%%%%%%%%%%%%%%%%%%%%%%%%%%%%%%%%
%%% DISCUSSION
%%%%%%%%%%%%%%%%%%%%%%%%%%%%%%%%%%%%%%%%%%%%%%%%%%%%%%%%%%%%%%%%%%%%%%%%%%%%%
\section*{Discussion}\label{sec6}
We put forward a novel scenario for the re-establishment of the epigenetic state of cells after DNA duplication. Key in this scenario is the formation of a liquid reaction container by polymer-assisted condensation (PAC) where the chromosome sections responsible for the condensate formation are simultaneously the sections that need to be restored. With only these sections residing inside the reaction container, enzymes in the condensate add marks mostly to nucleosomes at genomic positions that had been epigenetically marked in the previous cell generation. Using computer simulations, we demonstrated that this self-organised system is indeed capable of re-establishing the epigenetic state 50 times in a row, each time starting epigenetically diluted by a factor of two. This is consistent with the number of cell generations given by the Hayflick limit \cite{hayflick65}.

The proposed scenario 
exhibits remarkable robustness
to external perturbations. Both the changes of parameters and the complete interruption and subsequent restart of the process have little effect on the restoration of the epigenetic marks. 
Robustness is a general characteristic of living systems. However, the level of robustness required for this process to function effectively is particularly stringent:
(i) During DNA duplication half of the epigenetic marks are lost which effectively reduces the interaction strength between the heterochromatic sections and HP1 by a factor of two. PAC theory \cite{sommer22} shows that  the condensate has the remarkable property to be \textit{independent} of this interaction parameter (see also Eq.~\ref{eq:R3}). (ii) Experiments based on quantitative mass spectroscopy suggest that the restoration of histone lysine methylation is an extremely slow process taking about 20 hours \cite{xu12}. This process is interrupted by mitosis where the interphase chromosome structure is lost within minutes \cite{gibcus18}. After cell division, each chromosome needs to refold to an interphase state which allows the epigenetic restoration to continue. In fact, as demonstrated in Extended Data Fig.~\ref{fig:cell_cycle_stopped}, our simulations can be stopped at any time and restarted with a new initial configuration. PAC will then automatically self-organise the system into a structure that continues the process of epigenetic re-establishment.

It is by these two crucial properties of robustness that our study stands in contrast to the two previously proposed scenarios for explaining how cells restore their epigenetic state \cite{sandholtz20,owen23}.
These studies also simulate chromatin as a block copolymer with two types of monomers, representing nucleosomes belonging to eu- and heterochromatin. However, in these studies, the heterochromatic monomers attract each other either directly \cite{owen23} or via a bridging-type interaction \cite{sandholtz20}. In both cases, the simulation starts with a polymer that is allowed to move freely, leading to conformations where the heterochromatic sections are denser. In a second step, a radical approximation is introduced, namely the resulting polymer conformation is frozen. Only at this point, epigenetic marks are removed, without giving the polymer the possibility to adjust its configuration. A reaction scheme is then applied to the frozen polymer structure, bringing the labels preferentially back into the denser regions of the frozen polymer. When the reactions have reached a steady state, the cycle is finished and the new block copolymer is freed from its frozen state and can readjust its conformation.

Note that in this approach the memory of the epigenetic state is entirely stored inside the frozen configuration of the polymer. To cite from the abstract of Ref.~\cite{owen23}: ``In interphase, memory is held in the 3D structure of the genome...'' As the outcome of the reaction scheme is independent of the initial sequence, the reactions reconstruct the marks, even if the marks are removed altogether \cite{sandholtz20}. Furthermore, we note that these schemes with frozen configurations are not robust against a change of parameters. For instance, in Ref.~\cite{sandholtz20} one needs to carefully adjust the concentration of HP1 to avoid the model chromosome becoming entirely hetero- or euchromatic after a few generations. As shown in Fig.~3b of Ref.~\cite{owen23}, such a system is also extremely sensitive and unstable with respect to a change in the attraction strength between heterochromatic nucleosomes. As this is the parameter that effectively changes during the restoration of epigenetic marks, the system cannot stay within the stable range during most of the process. Limiting the number of available enzymes stops the runaway spreading of heterochromatic marks, Fig.~3e in Ref.~\cite{owen23}, but now the size of the heterochromatic domains directly reflects the number of available enzymes. That such a fine-tuning of the enzyme concentration occurs seems questionable, in addition to the necessity to freeze the chromatin conformation over the entire course of epigenetic re-establishment. In marked contrast, our scenario is robust with respect to changes in the parameters and allows for a dynamic rearrangement of  chromatin during the remethylation process. Our model shifts the focus from the role of particular conformational properties of chromatin to the collective behavior due to the formation of a biomolecular condensate. 

To conclude, we have proposed the first physically plausible scenario for cellular memory: epigenetic restoration based on PAC. Crucial elements of the scenario include: (1) a liquid reaction vessel as the only mesoscopic stable structure in this process, (2) a sharp boundary of the condensate constituting the physical realisation of the elusive boundary elements \cite{bannister01} and (3) enzymatic reactions with a preference for the condensate. Our scenario leads to a possibly unique combination of two types of robustness: independence of droplet properties from the polymer-protein interaction strength and independence from the initial chromosome conformation, allowing the process to be interrupted, as it likely occurs when mitosis interrupts the re-establishment process.

A possible weakness of our scenario is the requirement for cell cycle times to be chosen in a relatively small range to avoid systematic drift of heterochromatin domain boundaries. Small cycle times lead to a high probability of domain shrinkage, while large cycle times lead to likely domain expansion.
Closely related to this, our scenario does not suppress the spontaneous formation of a new heterochromatin domain within euchromatin. We believe that both problems can be resolved at the same time by introducing a more complex reaction scheme. In our current study, we simplified this system by assuming that there is only one reaction, the (full) methylation of a nucleosome. However, it is known that the H3K9me3 mark is produced in three separate steps \cite{cortini16}. Moreover, there is also a demethylation enzyme present \cite{cortini16}. This suggests that the accuracy of this process might be improved via kinetic proofreading \cite{hopfield74} where the methylation of a ``wrong'' nucleosome can be corrected. Also, the presence of both methylase and demethylase, with orthogonal selectivity with respect to the heterochromatin condensate, allows for a scenario where the activity of both enzymes cancel each other out at heterochromatin boundaries. This might make our process independent of the cell cycle time. We plan to implement these additional features into our model in a future study.

%%%%%%%%%%%%%%%%%%%%%%%%%%%%%%%%%%%%%%%%%%%%%%%%%%%%%%%%%%%%%%%%%%%%%%%%%%%%%
%%% ACKNOWLEDGEMENTS
%%%%%%%%%%%%%%%%%%%%%%%%%%%%%%%%%%%%%%%%%%%%%%%%%%%%%%%%%%%%%%%%%%%%%%%%%%%%%
\bmhead{Acknowledgements}
SM, ES, and HS were supported by the Deutsche Forschungsgemeinschaft (DFG, German Research Foundation) under Germany’s Excellence Strategy - EXC-2068 - 390729961.
\clearpage

\backmatter

%%%%%%%%%%%%%%%%%%%%%%%%%%%%%%%%%%%%%%%%%%%%%%%%%%%%%%%%%%%%%%%%%%%%%%%%%%%%%
%%% METHODS
%%%%%%%%%%%%%%%%%%%%%%%%%%%%%%%%%%%%%%%%%%%%%%%%%%%%%%%%%%%%%%%%%%%%%%%%%%%%%
\section*{Methods}\label{methods}
%%%%%%%%%%%%%%%%%%%%%%%%%%%%%%%%%%%%%%%%%%%%%%%%%%%%%%%%%%%%%%%%%%%%%%%%%%%%%
%%% COMPUTATIONAL MODEL
%%%%%%%%%%%%%%%%%%%%%%%%%%%%%%%%%%%%%%%%%%%%%%%%%%%%%%%%%%%%%%%%%%%%%%%%%%%%%

\noindent
\textbf{Computational model}\\
We carry out molecular dynamics (MD) simulations using a standard bead-spring model for the polymer which represents the model chromosome. We model HP1 molecules as 
free beads of the same diameter as the monomers in the background of an implicit solvent model. Pair interactions are Lennard-Jones potentials 
\begin{equation}\label{eq:LJ}
    U_{\mathrm{LJ}}(r) = 4\epsilon_{\mathrm{LJ}}\left[\left(\frac{b}{r}\right)^{12}-\left(\frac{b}{r}\right)^{6}\right.
    %\nonumber \\
    \left.-\left(\frac{b}{r_c}\right)^{12}+\left(\frac{b}{r_c}\right)^{6}\right].
\end{equation}
where $b=1$ is the bead diameter (in LJ units), $\epsilon_{\mathrm{LJ}}$ is the interaction strength and $r_c$ is the cutoff distance. Interactions between monomers are truncated at $r_c = 2^{1/6}$ and are thus purely repulsive; we use $\epsilon_{\mathrm{LJ}}=1$ throughout. The same holds for the interaction between HP1 and unmarked monomers.
There are attractive interactions between the HP1-beads, characterized by $\epsilon_{\mathrm{LJ}}\equiv\chi_{S}$, and between the epigenetically marked monomers and HP1, given by $\epsilon_{\mathrm{LJ}}\equiv\epsilon_{S}$; both interactions are truncated at a cut-off distance of $r_c=2.5$. In the absence of HP1, the polymer is thus in a good implicit solvent. We note that due to the hard-core repulsion, a minimum value of $\epsilon_{S}$, $\epsilon_{S0}\simeq0.6$, is necessary to realize a crossover from repulsion to adsorption of the HP1-beads with respect to the polymer chain \cite{galuschko19}. The bulk phase diagram of the LJ-system has been studied before \cite{watanabe12} and the critical point was found to be located at $\chi_{X}\simeq0.9$ and $c_{X}=0.32$. 
Throughout our simulations we use
$\chi_{S}=1.1$, 
which is
well above the critical point. The transition to the condensed phase is then located at an HP1 bulk concentration of about $c_b\simeq 0.075$. Simulations are carried out using the LAMMPS MD package \cite{plimpton95} in a cubic box of length $L=50$ using periodic boundary conditions.

The model chromosome is a block copolymer consisting of 350 monomers (= nucleosomes). 
We intialize simulations with an epigenetic sequence consisting of seven alternating blocks, four of which are unmarked (euchromatin blocks) and three are epigenetically marked (heterochromatin blocks). At the beginning of each cell generation, half of the marks are removed, as specified in more detail below. We then let the system equilibrate via an MD run. Equilibration is assessed by monitoring the radius of gyration of the heterochromatin domains which initially collapsed due to PAC.

We then carry out a sequence of MC cycles in which we attempt to methylate nucleosomes, followed by MD runs to equilibrate the system. The number of attempts per cycle is set to the total number of nucleosomes with missing epigenetic marks at the beginning of the cell cycle, which is 75. Randomly picked nucleosomes are then methylated with the probability $p_m$, given in Eq.~\ref{eq:meth_prb} of the main text. After finishing an MC cycle, we allow the system to respond to the changed sequence by performing an equilibration run of 50000 MD timesteps. As can be seen from Fig.~\ref{fig:PAC}(d), the droplet has exchanged most of its HP1 molecules by then. Furthermore, after this time, the monomers already show classical subdiffusion (see Extended Data Fig.~\ref{fig:polymer_msd} and the section on polymer dynamics below), indicating that the polymer is at least locally in equilibrium.
The total number of MD runs between methylation cycles sets the real timescale of a cell generation.

%%%%%%%%%%%%%%%%%%%%%%%%%%%%%%%%%%%%%%%%%%%%%%%%%%%%%%%%%%%%%%%%%%%%%%%%%%%%%
%%% QUANTIFY RE-ESTABLISHMENT
%%%%%%%%%%%%%%%%%%%%%%%%%%%%%%%%%%%%%%%%%%%%%%%%%%%%%%%%%%%%%%%%%%%%%%%%%%%%%

\vspace{.5cm}
\noindent
\textbf{Quantification of degree of re-establishment of epigenetic marks }\\
Here we describe the definitions of mismatches between the current epigenetic sequence and a target sequence that the system is attempting to reach through methylation reactions. These mismatches are used in Figs.~\ref{fig:one gen}(b) and \ref{fig:stretched_exp}. We first define the target sequence. We start from the 50 monomers long blocks of heterochromatin. After removing half of the epigenetic marks, blocks might have shrunk, as some of the nucleosomes at the boundaries have lost their marks. Our re-establishment scenario has no way to determine the exact locations of the original hetero-/euchromatin boundaries. We thus define the slightly smaller domains, including the outermost still methylated nucleosomes, as new heterochromatin domains. This sequence with all its defects inside heterochromatin domains closed is our target sequence. 
%\begin{figure}[ht]
%     \centering
%    \includegraphics[width=8cm]{final images/p_to_d.pdf}
%    \caption{Schematic showing the shift of domain boundary %after the restoration process. Ideal boundary length of 50 %nucleosomes is marked.}
%    \label{fig:ptod} 
%\end{figure}
We define now the heterochromatin mismatch as
\begin{equation}
    {\xi}_H=\frac{n_T- n_H(t)}{n_T}=\frac{\Delta n_H}{n_T}
\end{equation}
where $n_T$ is the total number of marked nucleosomes in the target configuration and $n_H(t)$ is the number of heterochromatin nucleosomes at time $t$. This quantity tracks how the defects of missing epigenetic marks are closed over time.

In addition, we have two more mismatches. These keep track of the methylation of ``wrong'' nucleosomes: the mismatch ${\xi}_b$ due to the growth of heterochromatin into euchromatin (the boundary growth) and the mismatch ${\xi}_e$ that results from the spontaneous formation of epigenetic marks on nucleosomes inside the euchromatin bulk. Specifically 
\begin{equation}
    {\xi}_b=\frac{\Delta n_b}{n_T} , {\;}{\;}{\;}{\;}      {\xi}_e=\frac{\Delta n_e}{n_T}  
\end{equation}
Here $\Delta n_b$, $\Delta n_e$ again count errors by comparing with the target configuration.

%%%%%%%%%%%%%%%%%%%%%%%%%%%%%%%%%%%%%%%%%%%%%%%%%%%%%%%%%%%%%%%%%%%%%%%%%%%%%
%%% SELECTIVITY
%%%%%%%%%%%%%%%%%%%%%%%%%%%%%%%%%%%%%%%%%%%%%%%%%%%%%%%%%%%%%%%%%%%%%%%%%%%%%
\vspace{.5cm}
\noindent
\textbf{Selectivity}\\
The proposed methylation scheme in the main text introduces two new parameters, $p_m^0$ and ${\epsilon}_m$, see Eq.~\ref{eq:meth_prb}. These two parameters determine the rates of the methylation reactions and can be tuned to achieve a selective, fast re-establishment of epigenetic marks. In order to find suitable parameters, we introduce a quantity, 
we call selectivity $\Lambda$, which measures how much more likely ``right'' nucleosomes are methylated compared to ``wrong'' ones. Specifically 

\begin{equation}
 \Lambda=\Big (\frac{1}{N_\textbf{S}}{\sum_{i \in \text{\textbf{S}}}P_m^i} \Big ) \Big / \Big ( \frac{1}{N_\textbf{B}}{\sum_{j \in \text{\textbf{B}}}P_m^j}\Big )
 \label{eq:selectivity}
\end{equation}
where $P_m^{x}$ is the total integrated methylation probability of nucleosome $x$, $P_m^x=\int dn{\;} w_x(n)p_m(n)$ with $p_m(n)$ given by Eq.~\ref{eq:meth_prb}; as in the main text, we set $p_m(n)$ to one whenever it exceeds this value. $w_x(n)$ is the histone number distribution around nucleosome $x$, determined from the MD simulations. A simpler approach would be to just use the mean HP1 numbers but since the methylation probability scales exponentially with the number of HP1 proteins, a slight fluctuation can make a large contribution to the methylation probability of the monomer. Finally, $\textbf{S}$ and $\textbf{B}$ describe two sets of boundary nucleosomes. $\textbf{S}$ contains all $N_\textbf{S}$ nucleosomes that sit at the ends of defects inside heterochromatin whereas $\textbf{B}$ is the set of $N_\textbf{B}$ euchromatic nucleosomes just at the border to heterochromatin blocks. Extended Data Fig.~\ref{fig:selectivity_variation}(a) indicates the $\textbf{S}$-monomers for defects of length one, two and three nucleosomes. The rational behind this definition of $\Lambda$ is that good selectivity is achieved when the $\textbf{S}$-monomers get methylated first, moving the segment boundary of defects quickly inwards, before $\textbf{B}$-monomers start to get methylated.
Large values of $\Lambda$ therefore lead to the desired separation of time scales.

%%%%%%%%%%%%%%%%%%%%%%%%%%%%%%%%%%%%%%%%%%%%%%%%%%%%%%%%%%%%%%%%%%%%%%%%%%%%%
%%% POLYMER DYNAMICS
%%%%%%%%%%%%%%%%%%%%%%%%%%%%%%%%%%%%%%%%%%%%%%%%%%%%%%%%%%%%%%%%%%%%%%%%%%%%%
\vspace{.5cm}
\noindent
\textbf{Polymer dynamics inside droplet}\\
Here we study the dynamics of the heterochromatin sections. Unlike in earlier studies \cite{sandholtz20,owen23}, our polymer is free to move during the re-establishment of the epigenetic state and is highly dynamic. We show this here by reporting the dynamics for both extremes, the fully and the half-methylated states, using two different methods. In the first approach, we report the mean squared displacement (MSD) of the heterochromatin sections with respect to the center of mass of the droplet:
\begin{equation}
    <{r_d}^2>= \frac{1}{N_M}\sum_{i=1}^{N_M} \{ (\Vec{\Gamma}_i(t)-\Vec{\Gamma}_{CM}(t))-(\Vec{\Gamma}_i(0)-\Vec{\Gamma}_{CM}(0))\}^2
\end{equation}
where the summation is over all $N_M$ monomers that belong to heterochromatin sections. $\Vec{\Gamma}_i(t)$ is the position of monomer $i$ and $\Vec{\Gamma}_{CM}(t)$ is the center of mass of the condensate, both at time $t$.
In Extended Data Fig.~\ref{fig:polymer_msd} the MSD is plotted for fully methylated and half-methylated heterochromatin. Both curves show subdiffusive behaviour with an exponent 1/2, compatible with Rouse dynamics, see Eq.~(5.126) in Ref.~\cite{schiessel}. The Rouse regime sets in earlier for the half-methylated polymer which, for any given time, also shows overall larger MSD values. For longer times, not shown here, the MSD converges to a finite value, set by the size of the condensate. At $t=50000$, the time between two MC sweeps in our cell cycle simulations, the polymer already shows subdiffusive dynamics, indicating that the monomers are locally in equilibrium at this time.

To gain a better insight into the dynamic conformational rearrangement of the polymer sections inside the condensate, we introduce the neighbourhood-neighbourhood correlation function $C_{NN}$ which tracks the changes in the neighbourhood of monomers by mapping each monomer's proximity inside the droplet to a binary state. Specifically
\begin{equation}
    C_{NN}(\Delta t)= 1/({N_M}^2-N_M) \sum_{i} \sum_{j} n_{ij}(t_0 +\Delta t)n_{ij}(t_0) 
\end{equation}
where $N_M$ is the number of monomers inside the heterochromatin sections and $n_{ij}(t)$ is an elements of the neighbourhood matrix at time $t$:
\vspace{.2cm}
\begin{equation}
\mathbb{N}=
\begin{pmatrix}
    0       & n_{12} & n_{13} & \dots & n_{1 N_M} \\
    n_{21}       & 0 & n_{23} & \dots & n_{2 N_M} \\
    \vdots          & \vdots & \vdots &\ddots&\vdots\\
    n_{N_M 1}       & n_{N_M 2} & n_{N_M 3} & \dots & 0
\end{pmatrix}.
\end{equation}
The value of $n_{ij}(t)$ tracks which monomers are in contact with a particular monomer as follows:
\begin{equation}
n_{ij}(t)=
\begin{cases}
    0 & \text{for} \quad i=j,\\
    1 & \text{if} \quad d_{ij}(t)< R_D,\\ 
    -1 & \text{if} \quad d_{ij}(t)\geq R_D.\\
\end{cases}
\end{equation}
Here $d_{ij}(t)$ is the distance between $i$th and $j$th monomer; $R_D$ is the radius of the droplet. This construction is shown schematically in Extended Data Fig.~\ref{fig:cnn}(a). Note that $\mathbb{N}$ is invariant under $\textbf{E}(n)$ group operations on the polymer as a whole. Therefore, any global rotation or translation of the polymer chain will not affect the correlation; only local rearrangements contribute to loss in correlation.

We have simulated two polymer chains with half and fully methylated heterochromatin blocks. The decay in $C_{NN}$ in the Extended Data Fig.~\ref{fig:cnn}(b) indicates significant rearrangements of the polymer with time. The local rearrangements are slower for the fully methylated case but in both cases the dramatic decay in $C_{NN}$ indicates that there is a very dynamic rearrangement of the polymer inside the droplet.

\clearpage
\bmhead{Extended data figures}
%%%%%%%%%%%%%%%%%%%%%%%%%%%%%%%%%%%%%%%%%%%%%%%%%%%%%%%%%%%%%%%%%%%%%%%%%%%%%
%%% APPENDICES
%%%%%%%%%%%%%%%%%%%%%%%%%%%%%%%%%%%%%%%%%%%%%%%%%%%%%%%%%%%%%%%%%%%%%%%%%%%%%
\begin{appendices}

%%%%%%%%%%%%%%%%%%%%%%%%%%%%%%%%%%%%%%%%%%%%%%%%%%%%%%%%%%%%%%%%%%%%%%%%%%%%%
%%% EXTENDED DATA FIGURES
%%%%%%%%%%%%%%%%%%%%%%%%%%%%%%%%%%%%%%%%%%%%%%%%%%%%%%%%%%%%%%%%%%%%%%%%%%%%%
\renewcommand{\figurename}{Extended Data Figure}

\begin{figure}[b!]
     %\centering
    \includegraphics[width=18cm]{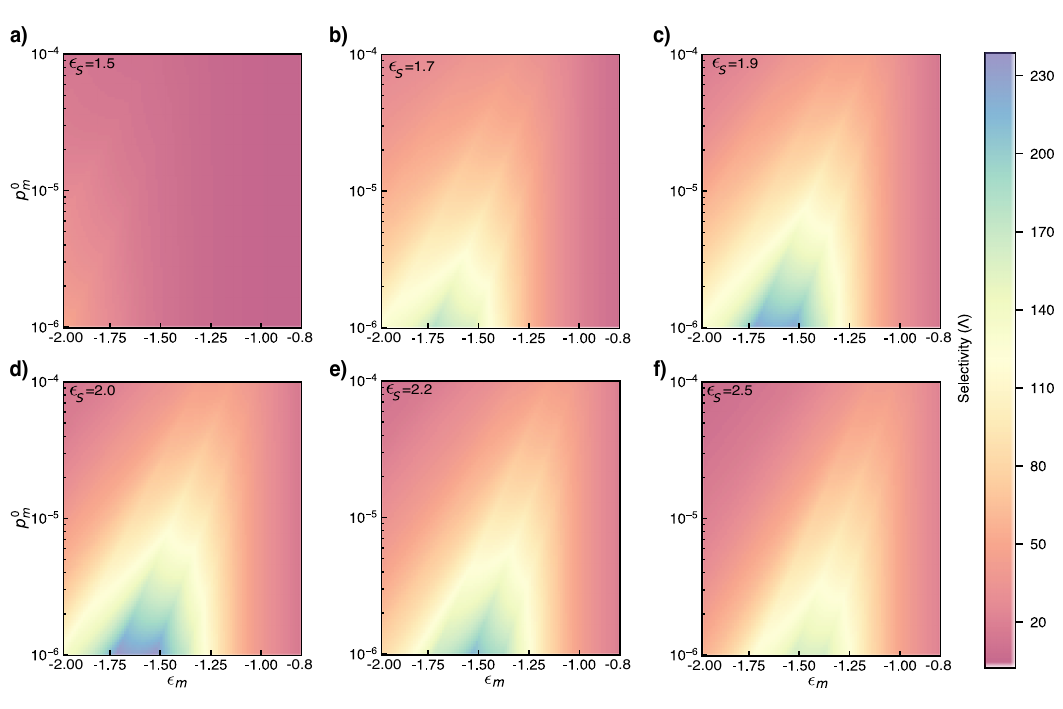}
    \caption{\textbf{Selectivity variation with ${\epsilon}_S$:} This panel shows maps of selectivity $\Lambda$ in the $p_m^0$--${\epsilon}_m$ parameter space for alternating heterochromatin sequences, as in Fig.~\ref{fig:selectivity}(b), but for different values of attraction strength between labeled nucleosomes and HP1. Specifically: (a) ${\epsilon}_S=1.5$, (b) ${\epsilon}_S=1.7$, (c) ${\epsilon}_S=1.9$, (d) ${\epsilon}_S=2.0$ (identical to Fig.~\ref{fig:selectivity}(b)), (e) ${\epsilon}_S=2.2$ and (f) ${\epsilon}_S=2.5$. Each case leads to high selectivity values, except for ${\epsilon}_S=1.5$. In this case ${\epsilon}_S/2$ is close to the mixed state, see Fig.~\ref{fig:PAC}(e)}
    \label{fig:selectivity_variation_eps_S} 
\end{figure}
\begin{figure}[h!]
     %\centering
    \includegraphics[width=18cm]{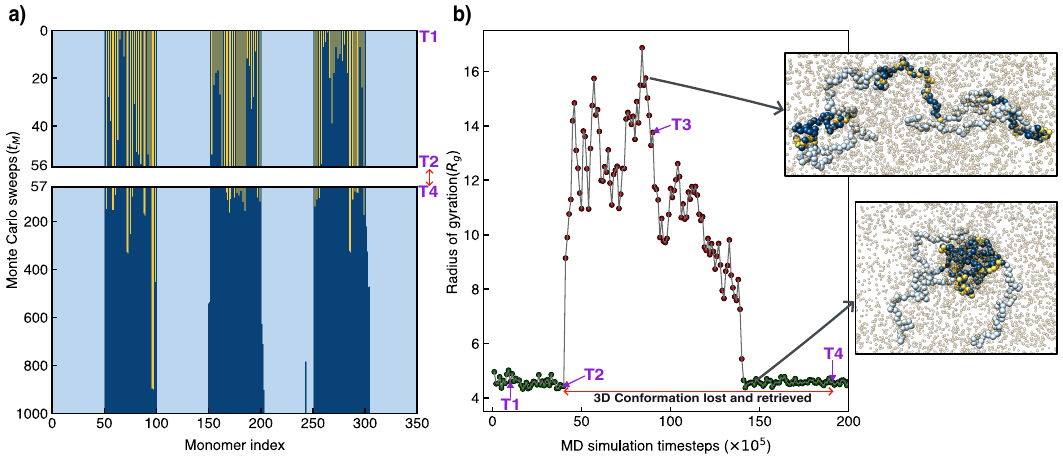}
    \caption{\textbf{Temporary disruption of condensate does not affect re-establishment of epigenetic marks:} (a) Development of epigenetic sequence from MC sweep 0, time $T1$, to MC sweep 56, time $T2$, and again, after disrupting the process, from MC sweep 57, time $T4$, onwards. The time $T2$ has been chosen when $75\%$ of the heterochromatic monomers were methylated. (b) Radius of gyration of heterochromatin sections. At time $T2$, the attraction between the marked monomers and HP1 is switched off, and the structure expands, see snapshot on top. After switching on the attraction again at time $T3$, the polymer collapses again into a micelle, see snapshot on bottom. The enzymatic reactions are switched off between time $T2$ and $T4$.}
    \label{fig:cell_cycle_stopped} 
\end{figure}
\begin{figure}[h!]
     %\centering
    \includegraphics[width=18cm]{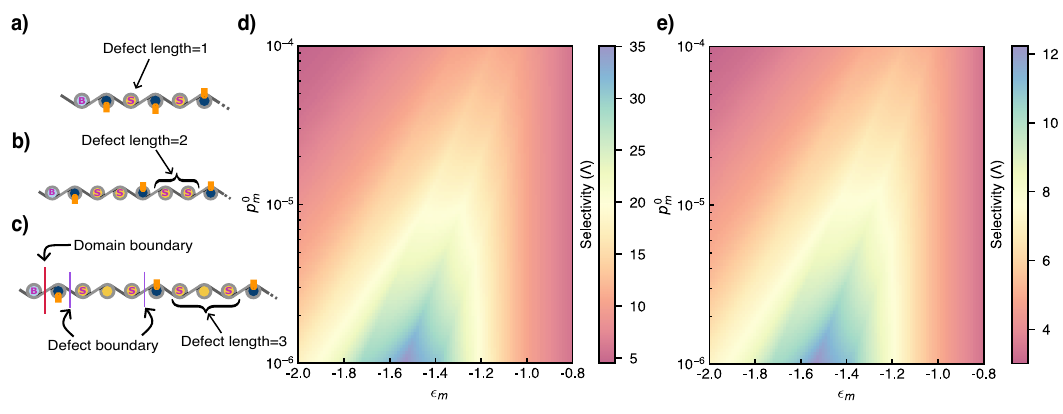}
    \caption{ \textbf{Selectivity variation with defect length:} Definition of monomer sets $\mathbf S$ and $\mathbf B$ in Eq.~\ref{eq:selectivity} for different defect lengths: (a) one, (b) two and (c) three nucleosomes. (d) Selectivity map for alternating sequence of blocks of two marked and two unmarked nucleosomes. (e) Same as (d) but for block size three. Selectivity values are strongly reduced compared to the case of block size one, Fig.~\ref{fig:selectivity}(b).}
    \label{fig:selectivity_variation} 
\end{figure}
\begin{figure}[h!]
     %\centering
    \includegraphics[width=18cm]{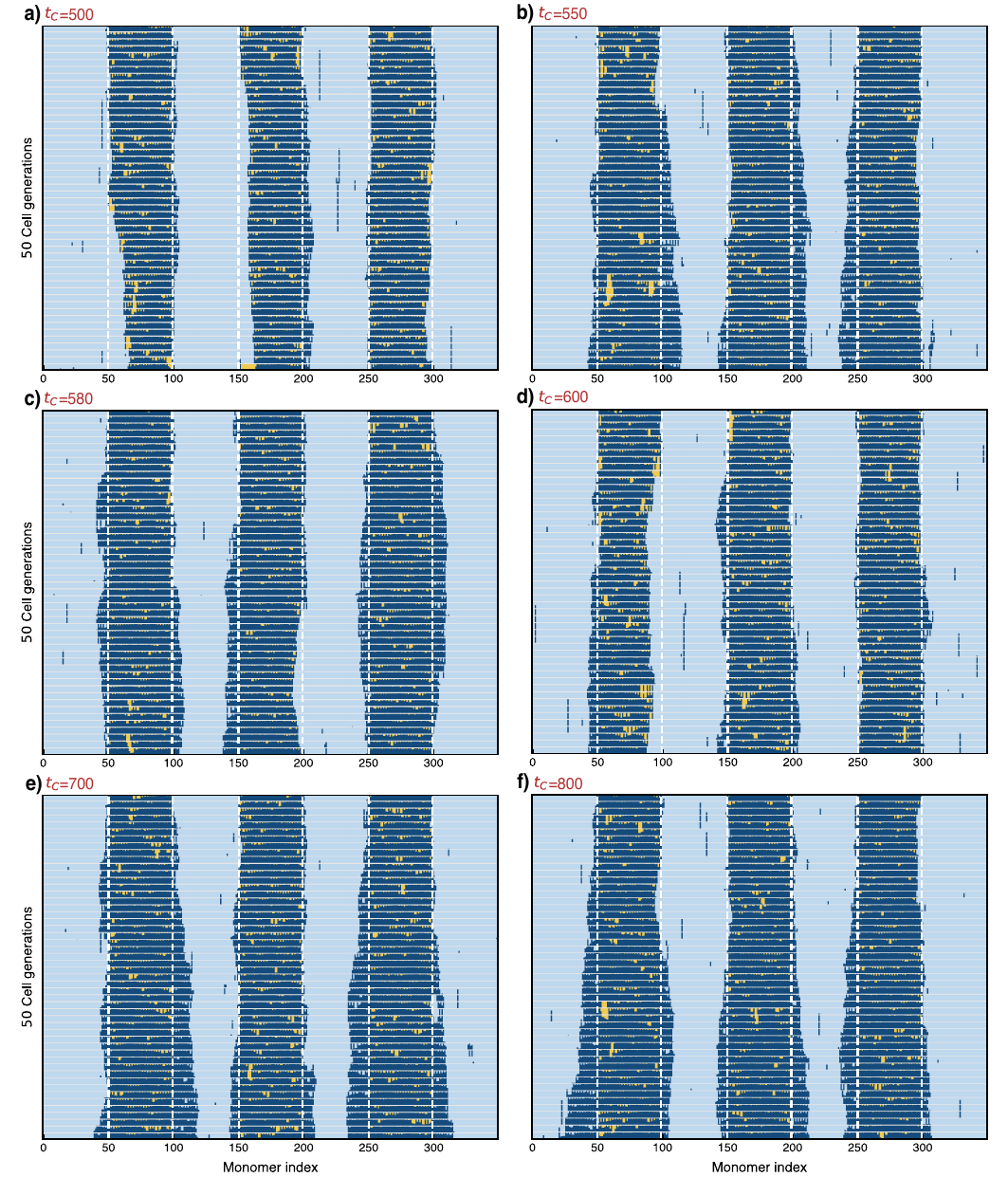}
    \caption{\textbf{Multiple cell-generation profiles for various cell cycle times:}
    Development of epigenetic sequences for 50 cell generations for different values of $t_C$, ranging from $t_C=500$ up to 800.}
    \label{fig:many} 
\end{figure}
\begin{figure}[h!]
     %\centering
    \includegraphics[width=18cm]{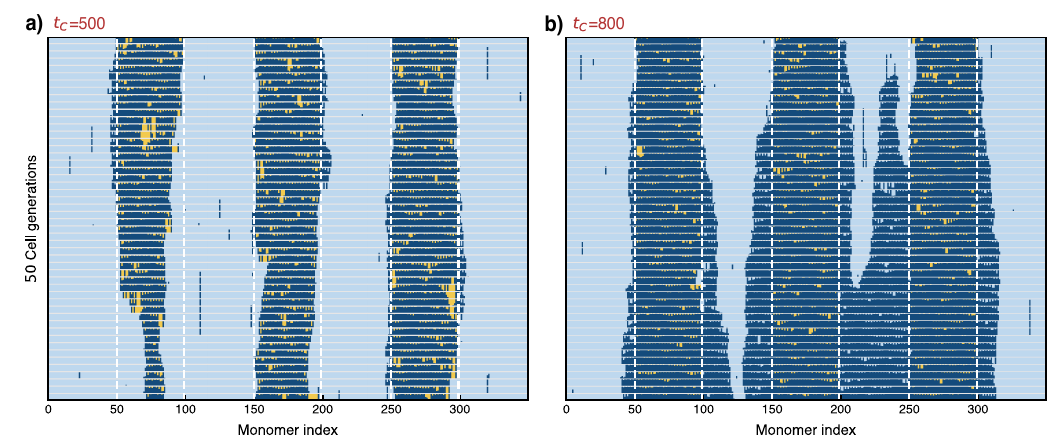}
    \caption{\textbf{Example of 50 cell-generation profiles with substantial changes in epigenetic sequence:}
    (a) A simulation run with a cell cycle time of 500 where the leftmost domain almost disappeared. (b) An example with cell cycle time 800 where a new domain formed within euchromatin that eventually glued the neighboring domains together.}
    \label{fig:error} 
\end{figure}
\begin{figure}[h!]
    \centering
    \includegraphics[width=8.8cm]{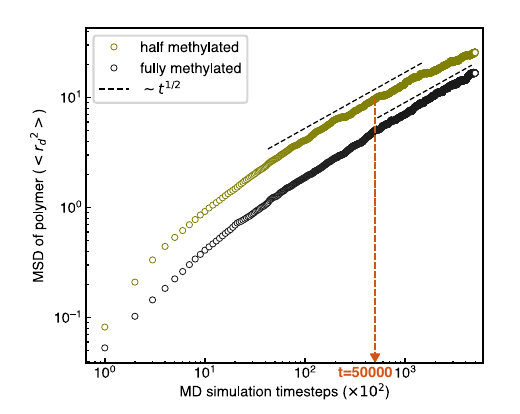}
    \caption{\textbf{Diffusion of heterochromatin inside droplet:}
    MSD of the monomers contained in the heterochromatin section as a function of the MD simulation time steps. Shown are curves for fully and half methylated heterochromatin. The lines indicate subdiffusive behaviour with exponent 1/2.}
    \label{fig:polymer_msd}   
\end{figure}
\begin{figure}[h!]
     \centering
    \includegraphics[width=18cm]{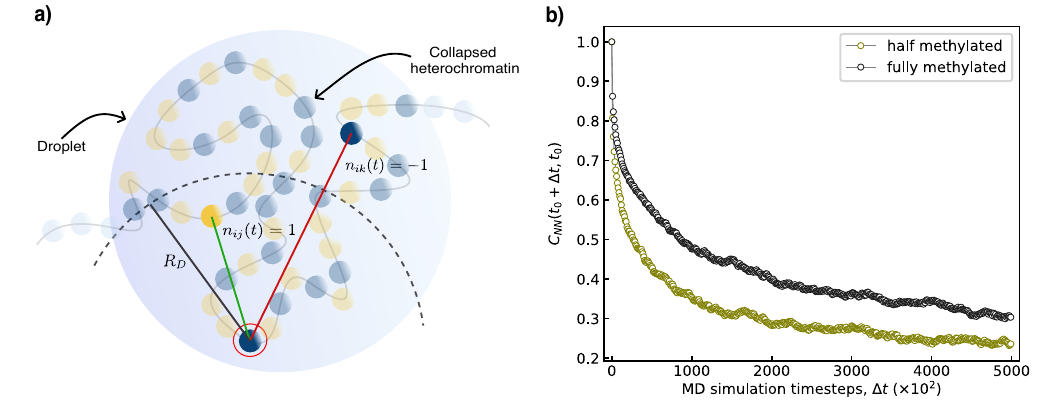}
    \caption{\textbf{Tracking the dynamic conformational rearrangement of heterochomatin:} (a) Schematics showing the mapping of the neighbours of a given monomer to a binary state. The neighbourhood radius is taken to be the radius of the droplet $R_D$. (b) Decay of neighbourhood-neighbourhood correlation with time for half and fully methylated chromosomes. 
    }
    \label{fig:cnn} 
\end{figure}

\clearpage
\end{appendices}
%\bmhead{Supplementary information}
%\begin{appendices}




%\end{appendices}


%%%%%%%%%%%%%%%%%%%%%%%%%%%%%%%%%%%%%%%%%%%%%%%%%%%%%%%%%%%%%%%%%%%%%%%%%%%%%
%%% REFERENCES
%%%%%%%%%%%%%%%%%%%%%%%%%%%%%%%%%%%%%%%%%%%%%%%%%%%%%%%%%%%%%%%%%%%%%%%%%%%%%
\bibliography{sn-bibliography}% common bib file

\end{document}